# Giant Nonlinear Hall Effect in Twisted Bilayer WTe$_2$

Zhihai He[1,2], and Hongming Weng[2,3,1*]

[1] *Songshan Lake Materials Laboratory, Dongguan, Guangdong 523808, China*

[2] *Beijing National Laboratory for Condensed Matter Physics, and Institute of Physics,*
*Chinese Academy of Sciences, Beijing 100190, China*

[3] *School of Physical Sciences, University of Chinese Academy of Sciences, Beijing 100190, China*

In a system with inversion symmetry broken, a second-order nonlinear Hall effect can survive even in the presence of time-reversal symmetry. In this work, we show that a giant nonlinear Hall effect can exist in twisted bilayer WTe$_2$ system. The Berry curvature dipole of twisted bilayer WTe$_2$ ($\theta = 29.4°$) can reach up to ~1400 Å, which is much larger than that in previously reported nonlinear Hall systems. In twisted bilayer WTe$_2$ system, there exists abundant band anticrossings and band inversions around the Fermi level, which brings a complicated distribution of Berry curvature, and leading to the nonlinear Hall signals exhibit dramatically oscillating behavior in this system. Its large amplitude and high tunability indicate that the twisted bilayer WTe$_2$ can be an excellent platform for studying the nonlinear Hall effect.

## I. INTRODUCTION

Over the last few decades, the study of various Hall effects has become an important topic in condensed-matter physics. For Hall effect or anomalous Hall effect, it is well known that an external magnetic field or magnetic dopants are necessary to break the time-reversal symmetry [1]. However, a recent theoretical study [2] predicted that a second-order nonlinear Hall effect can survive in the presence of time-reversal symmetry but with inversion symmetry broken. Subsequently, the experimental study on bilayer and multilayer WTe$_2$ [3-4] has validated this prediction in terms of the observed transverse nonlinear Hall-like current with a quadratic current-voltage characteristic. The nonlinear Hall effect arises from the non-vanishing dipole moment of Berry curvature in momentum space, i.e. Berry curvature dipole [2]. The theoretical predictions of the nonlinear Hall effect in a material can be achieved by calculating the Berry curvature dipole from band structure. Multiple materials [5-15] have been predicted to possess strong nonlinear Hall effect, such as the Weyl semimetals [5-8], the giant Rashba material bismuth tellurium iodine (BiTeI) under pressure [10], the monolayer WTe$_2$ and MoTe$_2$ with an external electric field [11], and the strained twisted bilayer graphene [12] or WSe$_2$ [13-14].

The experimental observations of the nonlinear Hall effect are mainly limited to the two-dimensional systems [3-4, 14-15]. It is well known that the twist angle between adjacent layers can be used as a new degree of freedom to modulate the electronic structures of two-dimensional system, which has attracted great attention since the recent discovery of the unconventional superconductivity and correlated state in the twisted bilayer graphene [16-17] and twisted bilayer of transition metal dichalcogenides [18]. Very recently, it was shown that the strained twisted bilayer graphene [12] and WSe$_2$ [13-14] present a large nonlinear Hall response. The experimental results demonstrate that there exist a significant nonlinear Hall effect in bilayer WTe$_2$ [3], it is interesting to know how the twist regulates the nonlinear Hall effect in this system.

In this work, we study the nonlinear Hall effect in twisted bilayer WTe$_2$ by using first-principle calculations combined with the semiclassical approach. Multiple twisted bilayer WTe$_2$ structures are constructed, and the twist angles range from 12° to 73°. It is found that the twisted bilayer WTe$_2$ has a more complicated band structure compared to the prefect bilayer system. We chose the twisted bilayer WTe$_2$ with twist angle $\theta = 29.4°$ as a typical system, and show that the Berry curvature dipole of twisted bilayer WTe$_2$ can be strongly enhanced.

## II. METHODS

The nonlinear Hall effect originates from the dipole moment of Berry curvature over the occupied states. In a system with time-reversal symmetry but broken inversion symmetry, when applied an oscillating electric field $E_c = \text{Re}\{\xi_c e^{i\omega t}\}$, a transverse response current $j_a = \text{Re}\{j_a^{(0)} + j_a^{(2\omega)}e^{2i\omega t}\}$ can be generated, where $j_a^{(0)} = \chi_{abc}\xi_b\xi_c^*$ and $j_a^{(2\omega)} = \chi_{abc}\xi_b\xi_c$ are rectified current and second-harmonic current, respectively. The nonlinear conductivity tensor $\chi_{abc}$ is associated with the Berry curvature dipole [2] as follows:

$$\chi_{abc} = -\varepsilon_{adc}\frac{e^3\tau}{2\hbar^2(1+i\omega\tau)}D_{bd}. \quad (1)$$

* Correspondence author: hmweng@iphy.ac.cn

Here, $D_{bd}$ is the Berry curvature dipole, $\varepsilon_{adc}$ is the Levi-Civita symbol and $\tau$ is the relaxation time. The Berry curvature dipole can be written as [2]

$$D_{bd} = \sum_n \int_k f_{nk}\frac{\partial\Omega_{nk,d}}{\partial k_b}, \quad (2)$$

where $f_{nk}$ is the Fermi-Dirac distribution function, and $\Omega_{nk,d}$ is the $d$ component of the Berry curvature. For a two-dimensional system, only the out-of-plane component ($z$) of Berry curvature is nonzero, which is given by [19]

$$\Omega_{nk,z} = 2i\sum_{m\neq n}\frac{\langle n|\partial H/\partial k_x|m\rangle\langle m|\partial H/\partial k_y|n\rangle}{(E_n - E_m)^2}, \quad (3)$$

where $E_n$ and $|n\rangle$ are eigenvalues and eigenwave functions, respectively.

The first-principle calculations are performed by using the Vienna ab initio simulation package (VASP) [20] with the projector-augmented wave potential method [21-23]. The exchange-correlation potential is described using the generalized gradient approximation (GGA) in the Perdew-Burke-Ernzerhof (PBE) form [24]. Spin-orbit coupling (SOC) is taken into account self-consistently. The energy cutoff of the plane-wave basis set is 350 eV. A vacuum region larger than 15 Å is applied to ensure no interaction between the slab and its image. In our optimization, all structures are fully relaxed until the force on each atom is less than 0.02 eV/Å. The Van der Waals interactions between the adjacent layers are taken into account by using zero damping DFT-D3 method of Grimme [25]. The maximally localized Wannier functions [26-28] for the $d$ orbitals of W and $p$ orbitals of Te are generated to compute the Berry curvature and Berry curvature dipole [29].

## III. RESULTS AND DISCUSSIONS

The bilayer WTe$_2$ exhibits an orthorhombic lattice, the optimized lattice constants are calculated to be $a_1$=3.447 Å and $a_2$=6.284 Å. We construct a series of twisted bilayer WTe$_2$ structures based on the method described in Ref. [30]. For simplicity, we start with the normal stacking of perfect bilayer WTe$_2$. As shown in Fig. 1, to construct the twisted bilayer WTe$_2$, a supercell lattice as bottom layer is built, where $b_1$ and $b_2$ are superlattice basis vectors generated by primitive basis vectors $a_1$ and $a_2$. Correspondingly, the supercell of top layer shares a mirror plane $\mathcal{M}_x$ with bottom layer. The twisted bilayer structure is formed by rotating the top layer around the original point by angle $\theta$ and translate the top layer by vector $b_2$, while the bottom layer is holding fixed. To ensure that the twisted bilayer structure is commensurate, it have to be made sure that $b_1$ is perpendicular to $b_2$. However, this condition is hard to fully meet because of the non-particularity of lattice constants. Approximately, we use a looser condition, for example, the angle between $b_1$ and $b_2$ ranging from 89° to 91° is acceptable. Let $b_1=ma_1+na_2$ and $b_2=pa_1+qa_2$, in which $m$, $n$, $p$ and $q$ are integers, then the twist angle satisfies $\theta=2\arctan(m|a_1|/n|a_2|)$. It is important to note that the twist angle will change slightly after structural relaxation.

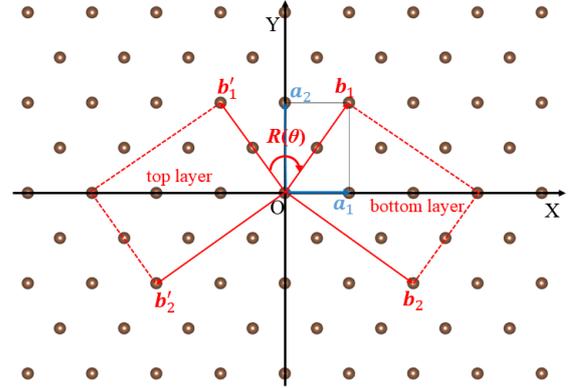

fig. 1. Schematics of the forming process of a twisted bilayer for orthorhombic lattice. $a_1$ and $a_2$ are basis vectors for primitive cell, while $b_1$ and $b_2$ are the supercell basis vectors of the twisted bilayer.

Table. 1 gives the optimized twist angle, formation energy and average interlayer distance for our constructed twisted bilayer WTe$_2$. Despite the twist angles ranging from 12° to 73°, these systems share pretty close formation energies and average interlayer distances. It is noted that the interlayer spacing in twisted bilayer WTe$_2$ is corrugated, and the average interlayer distance is slightly larger than that in perfect bilayer system (2.888 Å), suggesting a weakening of interlayer coupling.

Besides, the relatively smaller formation energy means that the twisted bilayer WTe2 may be formed in an efficient way.

Table. 1. The optimized twist angles $\theta$, formation energies $\Delta E$ (meV/Å$^2$) and average interlayer distances $\Delta Z$ (Å) for various twisted bilayer WTe2.

|  | (m, n) | (p, q) | $\theta$ | $\Delta E$ | $\Delta Z$ |
|---|---|---|---|---|---|
| W$_{344}$Te$_{688}$ | (1, 5) | (17, -1) | 12.3° | 6.37 | 3.296 |
| W$_{168}$Te$_{336}$ | (2, 5) | (8, -1) | 25.7° | 6.62 | 3.275 |
| W$_{60}$Te$_{120}$ | (1, 2) | (7, -1) | 29.4° | 6.67 | 3.284 |
| W$_{68}$Te$_{136}$ | (2, 3) | (5, -1) | 40.1° | 6.77 | 3.306 |
| W$_{116}$Te$_{232}$ | (5, 6) | (4, -1) | 49.1° | 6.55 | 3.307 |
| W$_{92}$Te$_{184}$ | (4, 3) | (5, -2) | 72.3° | 6.83 | 3.324 |

Compared to the twisted bilayer graphene, the twisted bilayer WTe2 exhibits much more complicated Moiré patterns since one monolayer WTe2 consists of one layer of W atoms and two layers of Te atoms. Figure 2(a) and 2(b) depict the top and side views of the optimized superlattice for twisted bilayer WTe2 with twist angle $\theta = 29.4°$ (the structures with other twist angles are given in the Fig. S1 in the Appendix). Such complicated Moiré superlattice leads to an intricate electronic structure in twisted bilayer WTe2. Take the system with twist angle $\theta = 29.4°$ as an example, there exist extensive band anticrossings and band inversions around the Fermi level, as shown in Fig. 2(c). With SOC being considered, the band structure becomes more intricate, which can be seen from Fig. 2(d). Similarly, the band structures with other twist angles also exhibit complicated band structures around the Fermi level, as shown in Fig. S2. The multiple bands cross or nearly cross in momentum space may bring about large gradient of Berry curvature around the band edges and result in strong nonlinear Hall response [2]. Next we focus on the twisted bilayer WTe2 with twist angle $\theta = 29.4°$, which possesses the smallest number of atoms in a superlattice, to calculate its Berry curvature dipole and estimate the nonlinear Hall effect.

The Berry curvature dipole $D_{xz}$ and $D_{yz}$ of twisted bilayer WTe2 ($\theta = 29.4°$) as a function of the chemical potential are shown in Fig. 3 (a). For a two-dimensional system, the Berry curvature dipole takes the unit of length. It is noted that in perfect bilayer WTe2 system, due to the presence of $\mathcal{M}_y$ mirror plane, only the $y$ component of Berry curvature dipole is nonzero. However, the introduction of twist can break this symmetry, so the $x$ component of Berry curvature dipole is also nonzero in the twisted bilayer

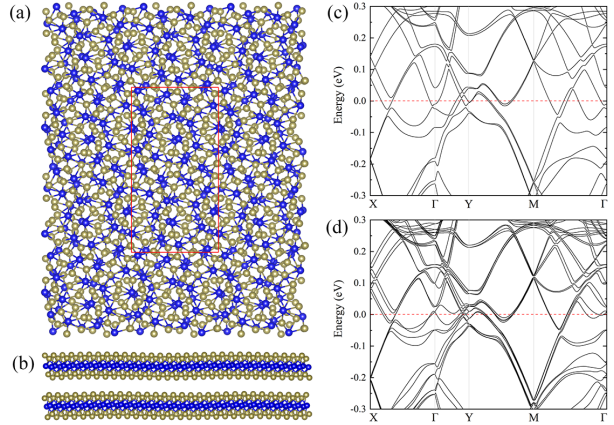

Fig. 2. (a), (b) The Moiré superlattice structure of twisted bilayer WTe2 ($\theta = 29.4°$) in top view and side view, respectively. The red rectangle indicates the supercell. The blue and yellow balls represent W and Te atoms, respectively. (c), (d) The band structure of twisted bilayer WTe2 ($\theta = 29.4°$) without and with SOC, respectively.

WTe2. It is clear that the Berry curvature dipole $D_{xz}$ and $D_{yz}$ exhibit drastic oscillating behavior near the Fermi level, and $D_{xz}$ and $D_{yz}$ can switch their signs dramatically within a very narrow energy region. On the other hand, we find that the magnitude of Berry curvature dipole in our considered twisted bilayer WTe2 is much larger than that in previous reports [3-4, 11-15]. For example, the peak of $D_{yz}$ locates near the Fermi level, which is calculated to be ~1400 Å. As a comparison, the Berry curvature dipole in monolayer or bilayer WTe2 [3, 11] is estimated to be in the order of 10 Å, and ~200 Å in strained twisted bilayer graphene [12].

To have a clearly understanding of the features of Berry curvature dipole in the twisted bilayer WTe2, we analyze the band structure and the distribution of Berry curvature. In general, the large Berry curvature appears at the band edge, as displayed in Fig. 3(b), where shows the band structure and Berry curvature $\Omega_{nk,z}$ along the $\Gamma - Y$ line. As can be seen from Fig. 3(b), the entanglement of multiple bands around the Fermi level causes a complicated distribution of Berry curvature, while the large Berry curvature dipole arises from the drastic change of Berry curvature in momentum space, as indicated by Eq. (2). We find that the peaks of Berry curvature dipole show in Fig. 3(a) mainly come from the contribution of the Berry curvature near the band edge shown in Fig. 3(b), for instance, the peaks of Berry curvature dipole originate from the band edges at corresponding energy

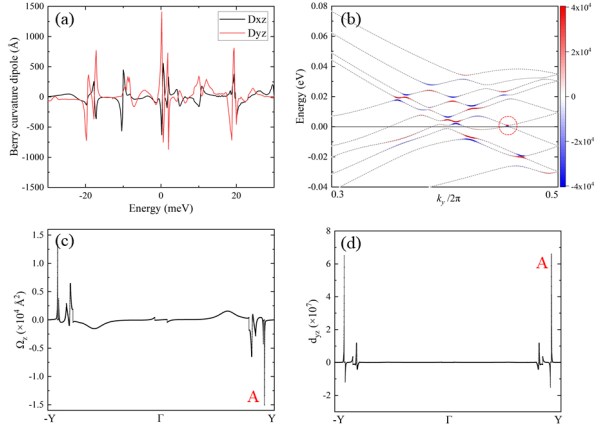

Fig. 3. (a) The Fermi energy dependence of Berry curvature dipole for twisted bilayer WTe$_2$ ($\theta$ = 29.4°). (b) Band structure and Berry curvature $\Omega_{nk,z}$ along Γ-Y line. The calculated Fermi level is set to be zero. (c) and (d) Berry curvature $\Omega_z$ and Berry curvature dipole density $d_{yz}$ along Y-Γ-Y line, respectively.

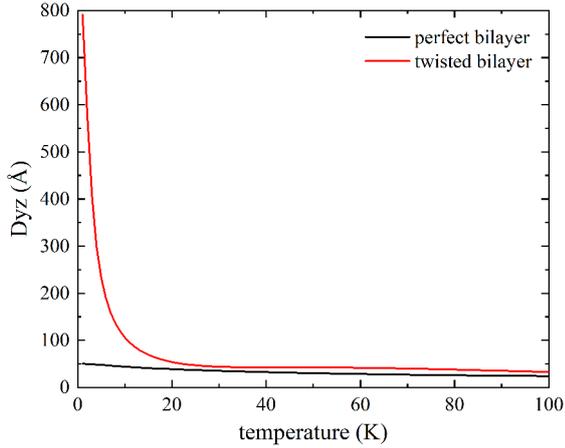

Fig. 4. The temperature dependence of Berry curvature dipole $D_{yz}$ for twisted bilayer WTe$_2$ ($\theta$ = 29.4°) and prefect bilayer WTe$_2$.

about -0.02, -0.01, 0.0 and 0.02 eV.

When the Fermi level is set to the charge neutral point, we plot the Berry curvature $\Omega_z$ and the corresponding density of Berry curvature dipole $d_{yz}$ along the Y − Γ − Y direction in Figs. 3(c) and 3(d), respectively. Here, the density of Berry curvature dipole in momentum space is defined as $d_{bd} = \sum_n f_{nk} \frac{\partial \Omega_{nk,d}}{\partial k_b}$. Clearly, the Berry curvature and $d_{yz}$ are mainly distributed in the region near the Y point, and the tremendous change of Berry curvature contributes large magnitudes to $d_{yz}$. Furthermore, it is found that the pronounced peak (marked by "A") shown in Figs. 3(c) and 3(d) indeed comes from the contribution of band edge indicated by the red circle in Fig. 3(b). As the Fermi level is shifted from the charge neutral point, the Fermi level go through multiple band anti-crossings in a narrow energy region, leading to the dramatic sign change of Berry curvature dipole around the charge neutral point.

As mentioned above, the Berry curvature dipole in twisted bilayer WTe$_2$ exhibits dramatically oscillating behavior near the charge neutral point. This indicates that the strong nonlinear Hall effects in our considered system may be observed at low temperatures. For clarity, we calculate the temperature dependence of Berry curvature dipole $D_{yz}$ for twisted bilayer WTe$_2$ ($\theta$ = 29.4°) and perfect bilayer WTe$_2$ according to $D_{yz}(T) = \int D_{yz}(E)(-\partial f/\partial E)dE$. Here, for simplicity, we fix the Fermi level to the charge neutral point. As shown in Fig. 4, the Berry curvature dipole $D_{yz}$ of perfect bilayer WTe$_2$ decreases as the temperature raises. This is qualitatively consistent with the experimental result in few-layer WTe$_2$ [4], where the nonlinear Hall response decreases monotonically with increasing temperature. When temperature is larger than 20 K, the Berry curvature dipole $D_{yz}$ for the twisted bilayer WTe$_2$ ($\theta$ = 29.4°) is slightly larger than that in perfect bilayer WTe$_2$. However, the Berry curvature dipole $D_{yz}$ for the twisted bilayer increases rapidly when temperature is less than 20 K. These results suggest that one could detect a very strong nonlinear Hall response at low temperatures in twisted bilayer WTe$_2$. On the other hand, the Berry curvature dipole in twisted bilayer WTe$_2$ is sensitively dependent on the Fermi level, its magnitude and sign can be switched dramatically within a very narrow energy region. The large magnitude and highly tunable characteristics of Berry curvature dipole in twisted bilayer WTe$_2$ provide an excellent platform to investigate the nonlinear Hall effect.

In this work, we focus on the nonlinear Hall effect in twisted bilayer WTe$_2$ with twist angle $\theta$ = 29.4°. Considering the large computational costs, the nonlinear Hall effect of other twist angle systems are not calculated. Nevertheless, the giant nonlinear Hall effect is expected to exist in other twist angle systems. For example, there also exist rich band crossings in the case of twist angles $\theta$ = 12.3° and $\theta$ = 25.7° (see Fig. S2).

## IV. CONCLUSIONS

In conclusion, we have predicted that a giant nonlinear Hall effect can exist in twisted bilayer WTe$_2$ system. We show that the twist can greatly change the band structure of bilayer WTe$_2$. There exist abundant band anticrossings and band inversions around the Fermi level in the twisted bilayer WTe$_2$, which brings a strong nonlinear Hall signal in this system. The Berry curvature dipole of twisted bilayer WTe$_2$ ($\theta$ = 29.4°) can reach up to ~1400 Å, much larger than that in previously reported nonlinear Hall systems. In addition, the nonlinear Hall effect in twisted bilayer WTe$_2$ exhibits dramatically oscillating behavior due to the complicated distribution of Berry curvature around the Fermi level. Our results show that the twisted bilayer WTe$_2$ can become an excellent platform to investigate the nonlinear Hall effect.

## ACKNOWLEDGMENTS

This work was supported by the National Natural Science Foundation of China (Grant No. 11925408, No. 12005153), the National Key Research and Development Program of China (Grant Nos. 2016YFA0300600 and 2018YFA0305700), the K. C. Wong Education Foundation (GJTD-2018-01), the Beijing Natural Science Foundation (Z180008), the Beijing Municipal Science and Technology Commission (No. Z191100007219013) and the Strategic Priority Research Program of Chinese Academy of Sciences (Grant No. XDB33000000). The computational resource is provided by the Platform for Data-Driven Computational Materials Discovery in Songshan Lake material Laboratory. The authors acknowledge the discussion with Prof. K. T. Law and Prof. N. Wang.

## APPENDIX

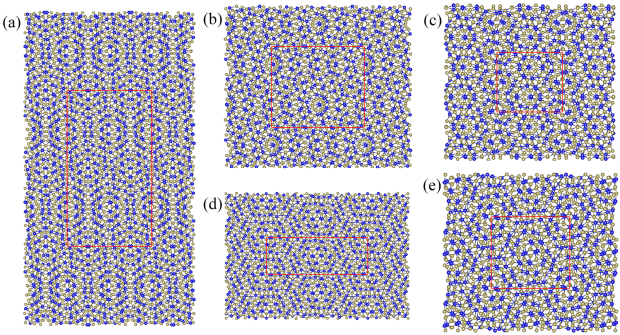

Fig. S1. The Moiré superlattice structures of twisted bilayer WTe$_2$ with twist angle 12.3° (a), 25.7° (b), 40.1° (c), 49.1° (d) and 72.3° (e). The red rectangle indicates the supercell.

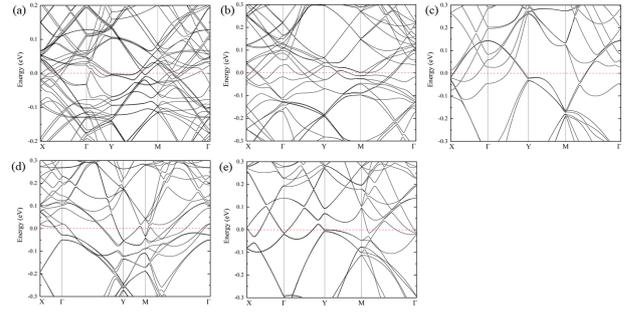

Fig. S2. Band structures (without SOC) of twisted bilayer WTe$_2$ with twist angle 12.3° (a), 25.7° (b), 40.1° (c), 49.1° (d) and 72.3° (e).